# Transmissive extreme ultraviolet metagrating

**Anna Kulter[1,*], Tiago Regio Crispim[1,2], Lorenz Weiss[1], Alexander Sagar Grossek[1], David J. Grafinger[1], Zsuzsanna Pápa[3], Judit Budai[3], Lázár Tóth[3], Péter Dombi[3,4], Rodolfo Previdi[5], Harald Plank[1], Andreas Hohenau[6], Martin Schultze[1], Marcus Ossiander[1,7,*]**

[1] *Graz University of Technology (Graz, 8010, Styria, Austria)*
[2] *Universidade do Minho (Braga, 4704-553, Norte, Portugal)*
[3] *ELI-ALPS, The Extreme Light Infrastructure ERIC (Szeged, 6728, Southern Great Plain, Hungary)*
[4] *HUN-REN Wigner Research Centre for Physics (Budapest, 1121, Central Hungary, Hungary)*
[5] *Institute of Science and Technology Austria (Klosterneuburg, 3400, Lower Austria, Austria)*
[6] *Institute of Physics, University of Graz (Graz, 8010, Styria, Austria)*
[7] *Harvard University (Cambridge, Massachusetts, USA, 02138)*
[*] *anna.karner@tugraz.at, marcus.ossiander@tugraz.at*

## Abstract

Extreme ultraviolet (EUV) radiation is a key tool for attosecond physics and lithography. However, strong material absorption limits the availability of transmissive optical elements at these wavelengths. Metaoptics exploit geometry to control the wavefront of transmitted light on the nanoscale and, due to their minimal thickness, promise to fill this gap. Here, we demonstrate the first EUV metaoptics for broadband applications: we design, fabricate, and experimentally investigate a blazed transmissive EUV metagrating and compare it with a focused-ion-beam-milled sawtooth-blazed grating serving as an in-situ reference. The metagrating achieves an angular dispersion of 0.04°/nm with a directionality (the ratio of the +1$^{st}$ and -1$^{st}$ diffraction order efficiency) of up to 5.8. The device shows phase-based operation up to 50 eV photon energy (down to 25 nm vacuum wavelength) and an octave-spanning bandwidth of 25 eV, doubling the previous spectral window addressable by metasurfaces. Comparing both gratings' performance reveals that, when accounting for fabrication constraints, EUV metasurfaces are competitive with free-form optics while offering scalability to large apertures and arbitrary phase profiles. Broadband transmissive operation removes the need for grazing incidence optics, defeating a major source of aberrations, and allows polarization-insensitive spectral analysis, enabling energy-resolved ultrafast spectroscopy in compact experimental configurations.

## Introduction

Only radiation in the extreme ultraviolet (EUV) spectrum provides the spectral bandwidth and the connected temporal resolution to resolve attosecond dynamics, and the wavelengths providing the Abbe limits required by modern lithography. State-of-the-art reflective optics enable powerful EUV focusing [1,2], spectrometers [3–5], and monochromators [6–11]. However, broadband performance demands grazing incidence operation which can be a source of aberrations in high-numerical aperture setups. This approach also requires large, costly, optics with complex surface shapes (e.g. ellipsoids) that are often unsuitable for compact experimental setups. Transmissive optical components would be an attractive alternative: they can be integrated in-line with the optical axis (without beam deviation), mitigating aberrations, enabling compact geometries, and making them suitable for a wide range of experimental configurations. However, strong material absorption and the lack of high-refractive index materials prohibitively limit the refractive power of realizable optics in this spectral region.

Metasurfaces use nanostructures with wavelength- and sub-wavelength-scale features and minimal thickness to control the spatial phase profile of transmitted light, predestining the concept for manipulating EUV light. A first metalens focusing (narrowband) 25 eV photon energy radiation to close to the diffraction limit [12] demonstrated the technology's viability. However, applying it in the future requires developing reliable nanofabrication protocols that can generate minuscule features, quantifying the influence of fabrication deviations, and examining its spectral performance.

Here, we do so by experimentally demonstrating the first blazed, transmissive, EUV metagrating and comparing its performance with an also first-of-its-kind sawtooth-blazed phase grating, as an in-situ, absolute reference. We introduce a simplified fabrication protocol, and find that, even in the presence of fabrication deviations, we achieve phase based performance over an ultrabroadband 25-eV spectral bandwidth. Although their phase profiles are simple, the abundant use of gratings and their defined, quantifiable function, as well as their ability to split properly manipulated and unaffected light, predestine them as an ultimate test for introducing metaoptics technology to broadband EUV light manipulation.

Existing EUV transmission gratings employ inversion-symmetric repeating structures such as lines or holes [13,14]. These binary phase and intensity gratings distribute light symmetrically into the positive and negative diffraction orders, i.e., they cannot preferentially direct the EUV intensity into a single diffraction order. This fact limits the diffraction efficiency and prevents directional (i.e., prism-like) light control. Other transmissive optical elements in the EUV are spectral filters [15,16], which block most of the EUV spectrum and transmit only a narrow wavelength band, or critical-angle transmission gratings which still rely on reflection [17].

As the prototypical diffractive optics, a grating (periodicity Λ) deflects normally incident light into positive and negative integer grating orders m emerging at vacuum wavelength-λ-dependent angles

$$\theta_m = \sin^{-1}(\frac{m\lambda}{\Lambda}). \tag{1}$$

and separates differently colored light in each order according to its angular dispersion $\mathrm{AD} = \frac{d\theta_m}{d\lambda}$.

Because Eq. (1)'s solutions appear in pairs ($\theta_m = -\theta_{-m}$), gratings usually redirect power into multiple (often unused) beams. Blazed gratings favor one diffraction order by imprinting a sawtooth-like phase profile $\Delta\varphi(x)$ along a spatial direction (here, x):

$$\Delta\varphi(x) = \frac{2\pi}{\lambda} x \sin(\theta_{+1}) \bmod 2\pi = 2\pi\,(\frac{x}{\Lambda} \bmod 1). \tag{2}$$

Each sawtooth locally acts like a prism that redirects the incident beam preferentially into the selected diffraction order's direction, while the periodicity enforced by the modulo operation provides the angular dispersion. Because an intensity-only mask cannot change the net momentum of an incident plane wave (deductible, e.g., from the angular spectrum of a real aperture [18], or from the generalized laws of refraction [19], or see Fig. 4 and its description), observing an enhanced diffraction efficiency into the +1st order and a suppressed diffraction efficiency into the −1st order reveals phase-based performance. To minimize the errors associated with absolute measurements, we introduce the diffraction efficiency ratio

$$\mathrm{DER} = \frac{\eta_{+1}}{\eta_{-1}} = \frac{I_{+1}}{I_{-1}}, \tag{3}$$

which is unambiguously extractable from experimentally measured light intensities ($I_{+1}, I_{-1}$) and modeled diffraction efficiencies ($\eta_{+1}, \eta_{-1}$).

**Results**

To create the phase profile in Eq. (2) using a metasurface, we follow the vacuum guiding approach [12] where holes (i.e., the absence of material, n = 1) etched into a substrate with a refractive index smaller than unity (n < 1) guide incident light and induce a hole diameter- and depth-dependent propagation phase shift upon transmission. Suitable substrate materials, i.e., such with a real part of the refractive index considerably smaller than one, and a sufficiently low imaginary part, are limited but available throughout the EUV spectrum [12,20]. For our design photon energy, 25 eV (vacuum wavelength 50 nm), e.g., silicon, aluminum, and beryllium, are suitable. Due to the availability of developed nanofabrication processes, we selected silicon for this work. The refractive index ñ = 0.77 + 0.02i of silicon (at the design wavelength of 50 nm, see Fig. S1 for silicon's refractive index) suggests that a 217-nm thin crystalline silicon layer transmits 34 % of the incident light and advances its phase by 2π compared to vacuum propagation, permitting emulating arbitrary transmission phase profiles. In practice, constraints on the minimally fabricable transverse feature sizes require a slightly longer propagation length for realizing a metasurface. Optimally, the holes (metaatoms) and their spacing in a metasurface should be smaller than the design wavelength (here, 50 nm) to avoid unwanted diffraction. To maintain fabricable feature sizes and aspect ratios, we chose a larger square tiling with a center-to-center distance between neighboring holes of 120 nm (unit cell size). This compromise does not change a metasurface's principal function but reduces the achievable diffraction efficiency (see below).

With the unit cell size set, we manufactured test samples containing different hole sizes by spin-coating a positive electron-beam resist (CSAR 62 [21]) onto a 500-nm thin crystalline silicon membrane (see Fig. S1 for its modeled transmission). The hole pattern was exposed in the resist using electron-beam lithography and subsequently transferred into the silicon by pseudo-Bosch inductively coupled plasma reactive ion etching (ICP-RIE, etch gas: $SF_6$, passivation gas: $C_4F_8$). In contrast to the conventional Bosch process, the pseudo-Bosch process etches and passivates simultaneously, suppressing scalloping and resulting in smoother hole sidewalls. We then sectioned the sample (see Fig. S2) using focused ion beam (FIB) milling to find the hole-diameter-$d_{hole}$-dependent etch depth $h_{hole} = 4\ d_{hole}$ - 70 nm for $d_{hole} \in$ [30 nm, 100 nm]. Using a relatively thick (500 nm) resist layer allows omitting a hard mask for etching, resulting in a high-yield single exposure-etch fabrication recipe (see Materials and methods and Fig. S3-S6).

We forward-designed our metagrating based on a metaatom library accounting for the smaller holes' limited etch rate. For this purpose, we simulated the far-field transmission phase of a periodic array of holes at the design wavelength as a function of the hole diameter and depth using rigorous coupled-wave analysis (RCWA, GRCWA [22], see Materials and methods). Varying the hole diameter from 35 nm (depth: ~70 nm) to 92 nm (depth: ~298 nm) achieves full $2\pi$ transmission phase coverage (exceeding that of the 220 nm thick membrane in [12]). We aimed to generate a first diffraction order m = +1 with a 0.04 °/nm angular dispersion suggesting a grating periodicity of Λ = 1.5 µm, however, experimentally found that using a grating periodicity divisible by the unit cell size (120 nm) increases resolving power. Therefore, we chose a 1.44 µm grating periodicity, discretized the spatially dependent grating phase profile (Eq. (2)), matched each phase with the correct hole diameter (see Fig. 1a), and manufactured a 100 µm × 100 µm-large metagrating. Fig. 1b shows a top-view scanning electron microscope (SEM) image of the final fabricated metagrating.

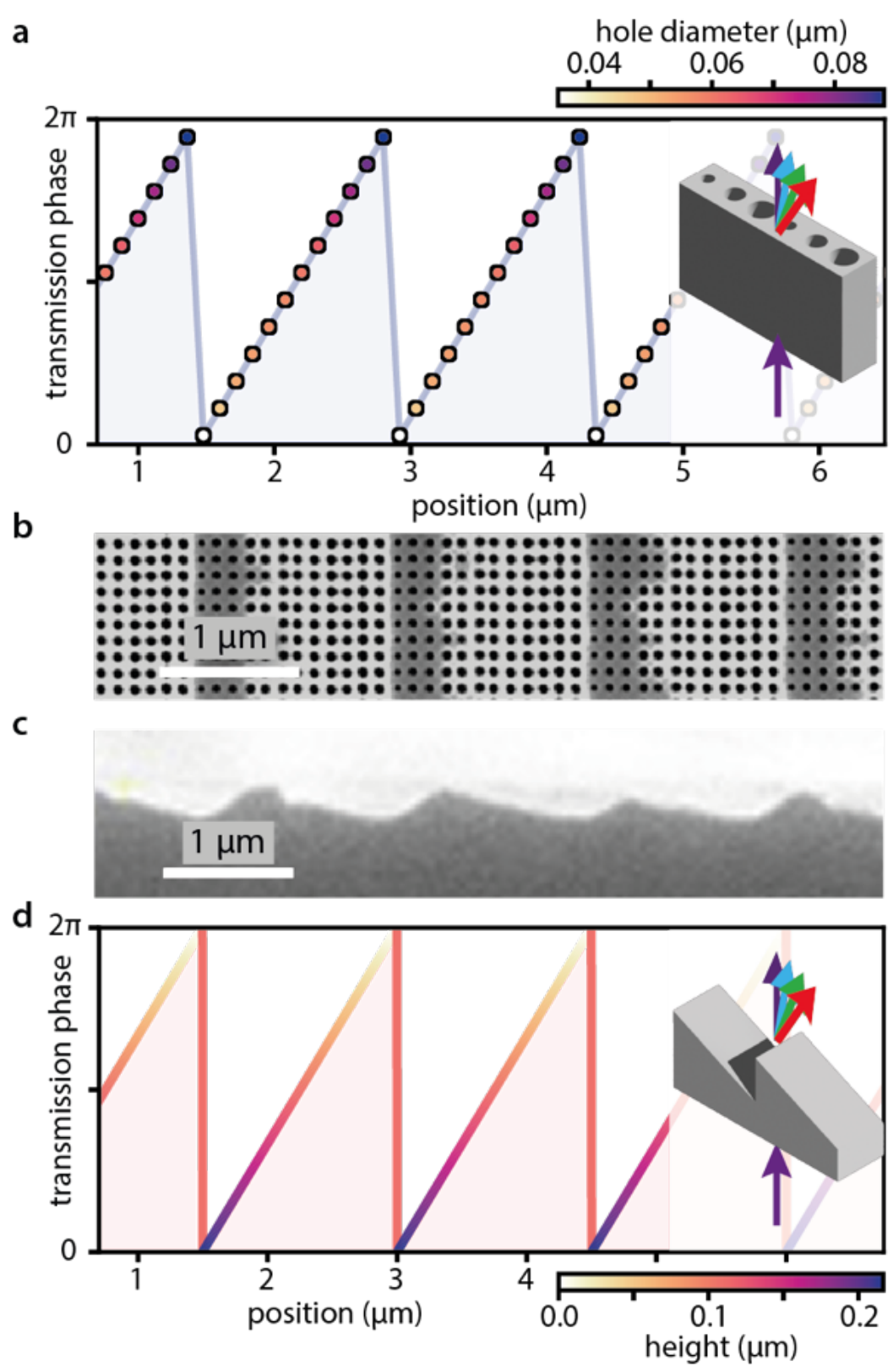

**Fig. 1: Fabrication of transmissive extreme ultraviolet (EUV) meta- and sawtooth gratings.** **a** Transmission phase profile of the metagrating. The positions of the holes are indicated by circles spaced 120 nm apart, with the color representing the hole diameter. Inset: 3D visualization of metagrating with incoming (bottom) and outgoing (top) beams. **b** Scanning electron microscope (SEM) image (top view) of the metagrating. **c** SEM image (side view) of a twin of the sawtooth grating on a silicon wafer. **d** Transmission phase profile of the sawtooth grating. The line color indicates the height of the grating teeth. Inset: 3D visualization of the sawtooth grating with incoming (bottom) and outgoing (top) beams.

As a reference, we also designed a conventional, 100 µm × 100 µm large, blazed grating by FIB milling an up to 220-nm deep sawtooth pattern repeating every 1500 nm, i.e., with a phase slope of 2π/1500 nm (see Materials and methods). Fig. 1d shows its designed transmission phase profile and sawtooth depth, and Fig. 1c shows a SEM image of the manufactured sawtooth cross-section. Note that, in this design, the tall side of a silicon sawtooth imprints the smallest phase shift on the incident EUV light (silicon refractive index $n < 1$).

To evaluate both gratings' performance, we generated high-harmonic radiation in argon [23–25] using ~160-fs long, s-polarized laser pulses centered at 1030 nm. A grazing-incidence toroidal mirror relay-imaged the source, producing an approximately 40-µm large focus (waist diameter at intensity full-width at half-maximum (FWHM)). This focus fits into the 100 µm device size while illuminating many grating periods. Because the devices operate in transmission, they can be positioned directly in the focus without deviating the zeroth diffraction order. A scientific CMOS camera placed 12.7 cm after the gratings recorded the diffraction patterns. Owing to the large Rayleigh range of the EUV beam, the spot size on the camera remained similar to that at the focus (~45 µm). Fig. 2a shows a schematic of the experimental setup.

To quantify the gratings' performance, Figs. 2c, d compare the $\pm 1^{st}$ order diffraction patterns generated by the metagrating and the sawtooth grating (lineouts integrated over 10-pixels along the y-direction of the camera image in Figs. 2b, S7). Gray lines indicate the expected positions of the odd high harmonics generated by the driving laser. Because argon and silicon are opaque below 25 eV photon energy (see Fig. S8), we begin our discussion slightly above the design photon energy, at 27 eV.

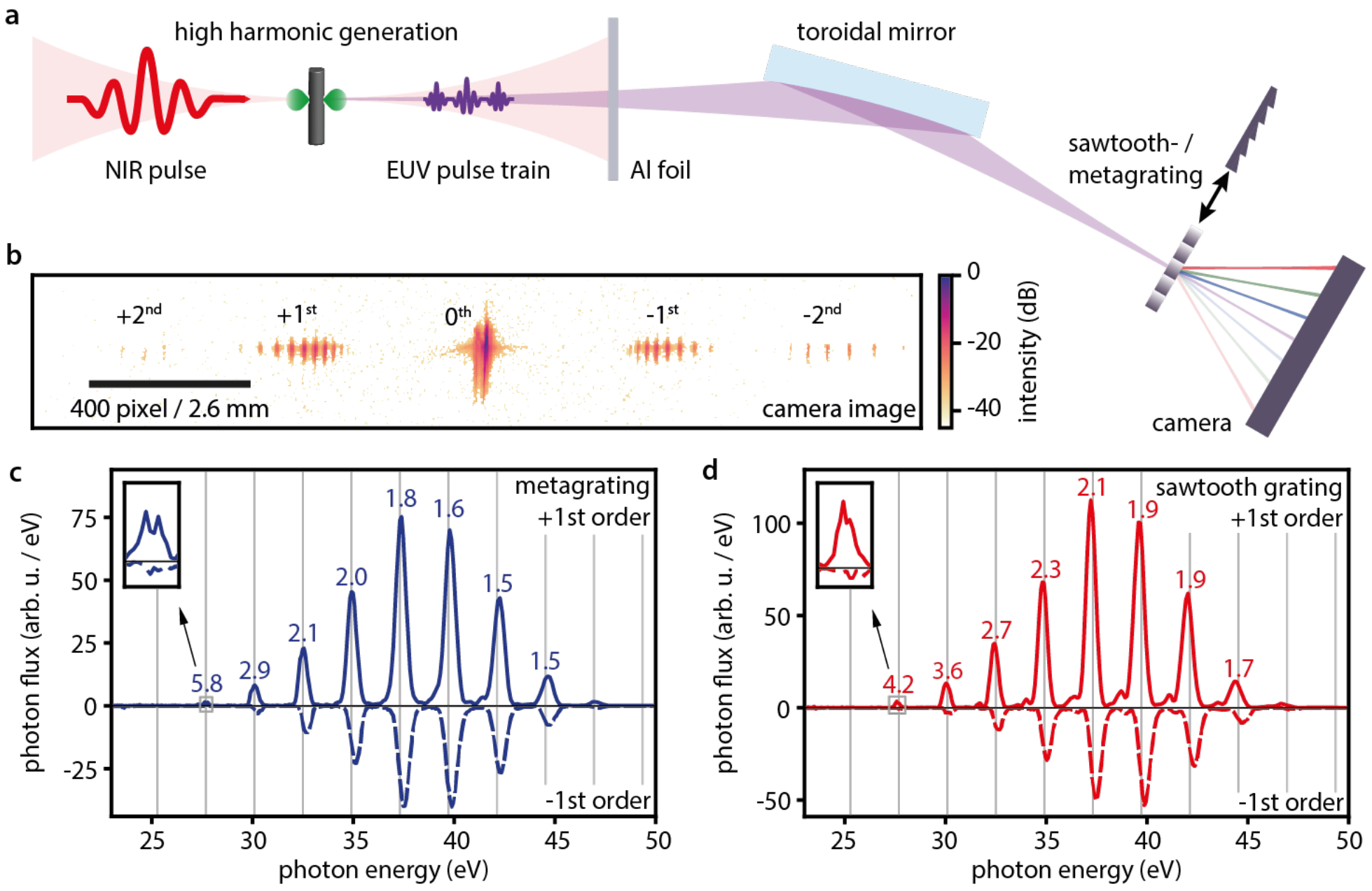

**Fig. 2: Experimental transmissive EUV grating characterization.** **a** An intense near-infrared (NIR) laser pulse generates an attosecond pulse train via high harmonic generation (HHG) in argon. A thin aluminum (Al) foil blocks the driving light and transmits the generated EUV radiation. The beam is then focused onto the gratings using a grazing-incidence toroidal mirror, and the resulting diffraction pattern is recorded with a CMOS camera. **b** Diffraction pattern of the metagrating on the CMOS camera. The numbers indicate the diffraction orders. Line-outs through the diffraction patterns of **c** the metagrating (blue lines) and **d** the sawtooth grating (red lines). The positive y-axes denote the +1$^{st}$ diffraction orders (solid lines), while the negative y-axes show the −1$^{st}$ diffraction order (dashed lines). The gray lines indicate the odd harmonics of the driving laser photon energy, and the numbers give the diffraction efficiency ratio (DER) at these energies. Lineouts extracted via a horizontal line profile (averaged over 10 pixels vertically) in the camera images shown in Fig. S7.

Both gratings suppress the −1$^{st}$ order and enhance the +1$^{st}$ order, demonstrating that they act as phase optics and achieve the intended blazed phase profile. At 27 eV, closest to the design photon energy, the metagrating's +1$^{st}$ diffraction order intensity is DER = 5.8 (see Eq. (3)) times stronger than the −1$^{st}$ diffraction order; the sawtooth grating achieves a ±1$^{st}$ order ratio of 4.2. In addition to the first diffraction order, both gratings exhibit second-order diffraction peaks at a larger distance from the undiffracted beam (see Fig. 2b). The metagrating provides a stronger suppression of the +2$^{nd}$ order, with a relative diffraction efficiency +1$^{st}$ / +2$^{nd}$ order ratio of 110 compared to the sawtooth grating's +1$^{st}$ / +2$^{nd}$ order ratio of 40 (averaged between 27 and 47 eV). In the sawtooth grating spectrum, small ghost peaks appear between the main high harmonic generation (HHG) peaks. We attribute these to slight variations in the grating period stemming from the time-intensive FIB milling process. In contrast, these features are absent in the metagrating spectrum, likely due to the higher positional accuracy of the holes achieved in the lithography process.

We determine the spectrally averaged diffraction efficiency by integrating the $+1^{st}$ order intensity between 27 and 47 eV and comparing it with the zeroth order intensity measured through an unpatterned spot of the silicon membrane. Due to the shape of our HHG spectrum, this performance is most indicative for the gratings' performance at 37 eV photon energy and a lower bound for the performance at smaller photon energies. While the metagrating provides a stronger suppression of unwanted diffraction orders near the design wavelength, the sawtooth grating exhibits a higher diffraction efficiency: it directs 15% of the total transmitted intensity into the $+1^{st}$ diffraction order, whereas the metagrating reaches 12.5%: because the metagrating's hole spacing of 120 nm exceeds the wavelength, it introduces a secondary grating period that redistributes part of the optical power to high diffraction angles – a tradeoff we accepted for manufacturability.

**Discussion**

To interpret these numbers, we simulate both gratings' performance using finite-difference time-domain modeling (FDTD, tidy3d, see Materials and methods). Optics performance in the EUV spectrum is influenced even by small fabrication deviations. To account for such, Fig. 3 presents modeled performance for the designed meta- and sawtooth gratings, as well as a metagrating whose smallest holes were enlarged by 20 nm during etching (worst case estimated from Fig. 1b) and a sawtooth grating that experienced redeposition of material during focused ion beam milling (shape estimated from Fig. 1c). Perfect gratings could achieve DERs beyond 60 at 27 eV photon energy (Fig. 3b, d), mainly by strongly suppressing diffraction into the $-1^{st}$ diffraction order (dark dashed lines in Fig. 3a, c). However, the suppression and DER drop when including fabrication deviations (lighter lines in Fig. 3). Fabrication deviations affect the sawtooth grating's modeled suppression decrease and +1$^{st}$ order diffraction efficiency more, explaining the gratings' comparable experimental performance. Because our devices operate in transmission and with normally incident light, they circumvent the polarization-dependence inherent to grazing-incidence reflection, and the metagrating's modeled +1$^{st}$ order diffraction efficiency differs by less than 1 % for s- and p-polarized light (Fig. S9). Its almost polarization-independent response predestines the device for detecting minuscule polarization changes exploited, e.g., in extreme-ultraviolet circular dichroism measurements [26–28].

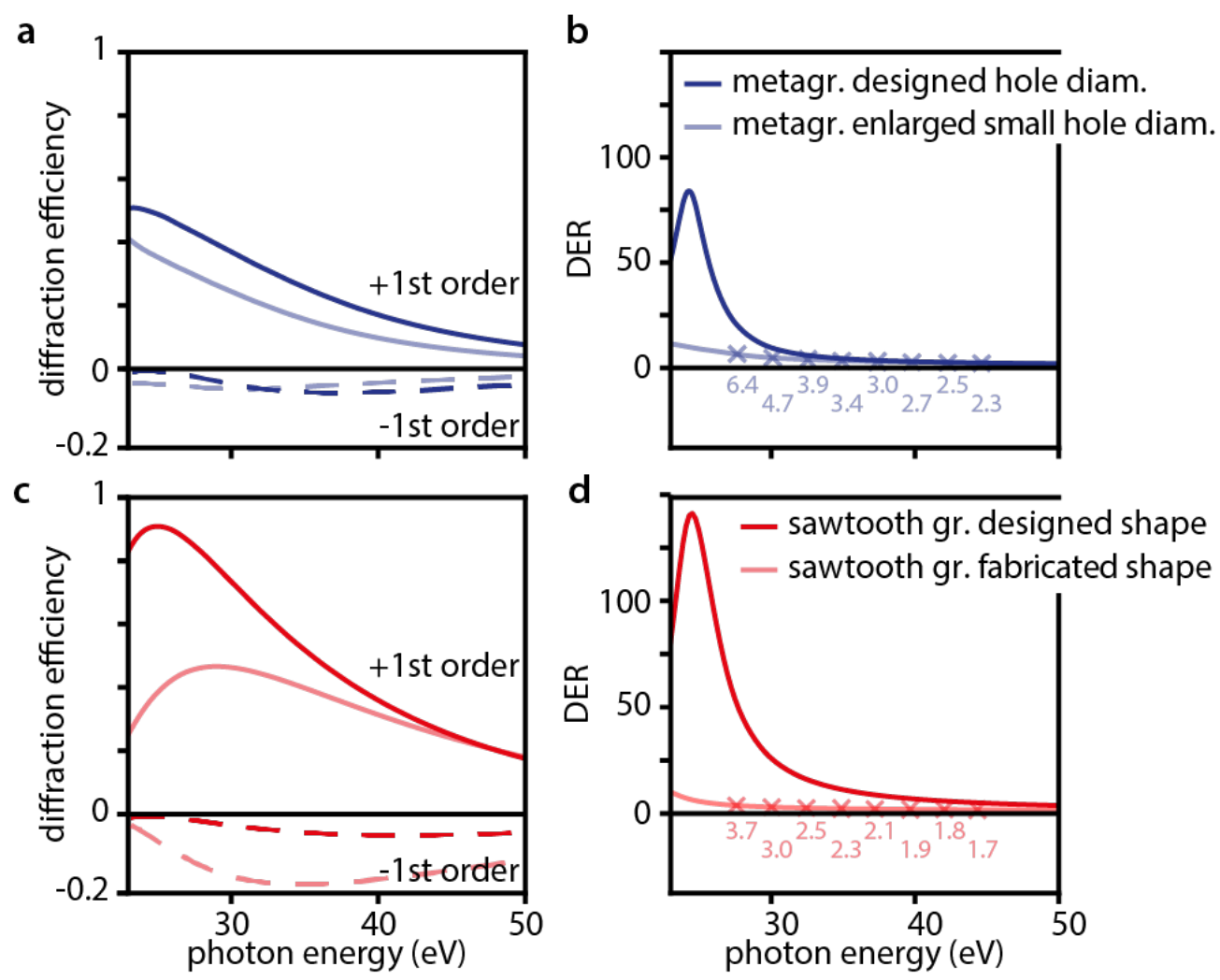

**Fig. 3 Modeled photon energy dependent diffraction efficiencies.** **a** Diffraction efficiency into the +1$^{st}$ (-1$^{st}$) diffraction order (relative to the full transmitted intensity) for a true-to-design metagrating (dark solid (dashed) blue curve) and a metagrating in which the smallest holes are enlarged by 20 nm (light solid (dashed) blue curve). **b** Metagrating diffraction efficiency ratio (DER). Crosses mark the high harmonic photon energies and numbers give DER values. **c** Sawtooth grating +1$^{st}$ (-1$^{st}$) order diffraction efficiency for a true-to-design grating (dark solid (dashed) red curve) and the fabricated sawtooth grating (light solid (dashed) red curve). **d** Sawtooth grating DER.

As we find evidence of phase-based behavior throughout the observed spectral region in the experiment and the modeling, we compressed the infrared driving laser pulse to approximately 10 fs to increase the peak intensity, extend the HHG spectrum to higher photon energies, and investigate the ultrabroadband performance of the gratings (Fig. 4a, b). The DER decreases with increasing deviation from the design photon energy, however, although the gratings were designed for a central photon energy of 25 eV, they remain functional up to at least 70 eV (our HHG cut-off energy). For the metagrating, the DER decreases from ~1.8 at 50 eV to 1.3 at 65 eV, whereas the sawtooth grating exhibits lower values over the same range (1.5 at 50 eV down to 1.0 at 65 eV). These results are in good agreement with our simulations, in which the metagrating suppresses the -1$^{st}$ diffraction order about 1.3 times better than the sawtooth grating at 50 eV and 1.2 times better at 65 eV (see Fig. S10).

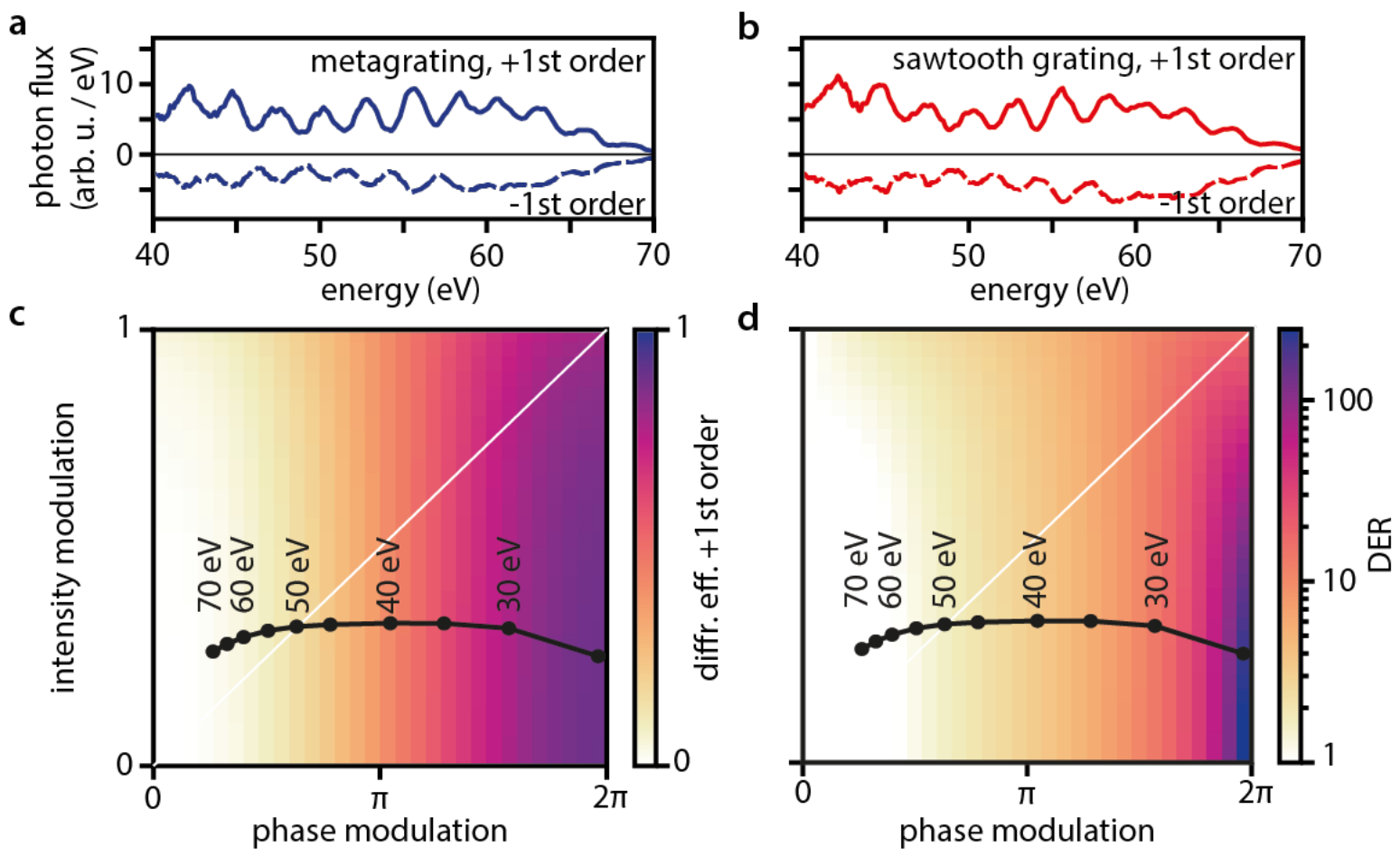

**Fig. 4: Experimental and modeled ultrabroadband performance.** Diffraction patterns at high photon energies of **a** the metagrating and **b** the sawtooth grating. The positive y-axes denote the +1$^{st}$ diffraction orders (solid lines) and the negative y-axes show the -1$^{st}$ diffraction order (dashed lines). **c** Modeled diffraction efficiency into the +1$^{st}$ diffraction order and **d** diffraction efficiency ratio (DER) of a flat mask as a function of its phase and intensity modulation depth. The markers show the phase and intensity difference of light transmitted through the smallest (35 nm diameter) and largest (92 nm diameter) holes in the metagrating design. In **c** and **d**, the area below the diagonal (white line) indicates primarily the behavior of a phase modulation grating, while above the diagonal, intensity modulation dominates.

To understand the broadband phase-based performance of the metagrating, we modeled diffraction by a flat intensity and phase mask (see Materials and methods). Figs. 4c, d show the expected diffraction efficiency into the +1$^{st}$ order and the DER as functions of the mask's phase and intensity modulation depth. Visibly, only phase gratings can achieve DER, which coincides with a higher total +1$^{st}$ order diffraction efficiency. The markers in Figs. 4c, d indicate the transmission phase and intensity modulation capability of our metaatoms for different photon energies (the phase and intensity difference photons experience when transmitting though the smallest (35 nm diameter) and largest (92 nm diameter) holes, see Fig. S11). The area below the diagonal (white line) represents mainly the behavior of a phase modulation grating, while the area above the diagonal represents primarily an intensity grating. The model reveals that the metagrating operates predominantly as a phase grating from 25 eV to 50 eV photon energy, i.e., it implements the desired blaze profile over a 25 eV absolute photon energy bandwidth (6 PHz frequency bandwidth), and a relative bandwidth of 0.66. Above 50 eV, the incoming wave is primarily modulated through absorption, however, the remaining phase modulation still maintains a larger than unity DER.

The observed ultrabroadband phase operation suggests optimizing the metagrating design for even broader operation. However, the current state of nanofabrication impedes active group-delay and group-delay-dispersion-control based on complex pillar shapes [29,30] in the EUV. Nonetheless, even simple adaptions can achieve valuable improvements: Figs. 4c, d suggest that increasing the phase modulation improves performance at higher photon energy. Therefore, we model a metagrating that was etched for longer, such that all holes deepen by 50 nm (while maintaining the hole diameters of our initial design). The results in Fig. 5 are convincing: for the price of only a slight performance decrease at 25 eV photon energy, an overetched sample can maintain more than 20 % diffraction efficiency into the +1$^{st}$ diffraction order and a DER exceeding 4 across the 25 eV to 50 eV photon energy range.

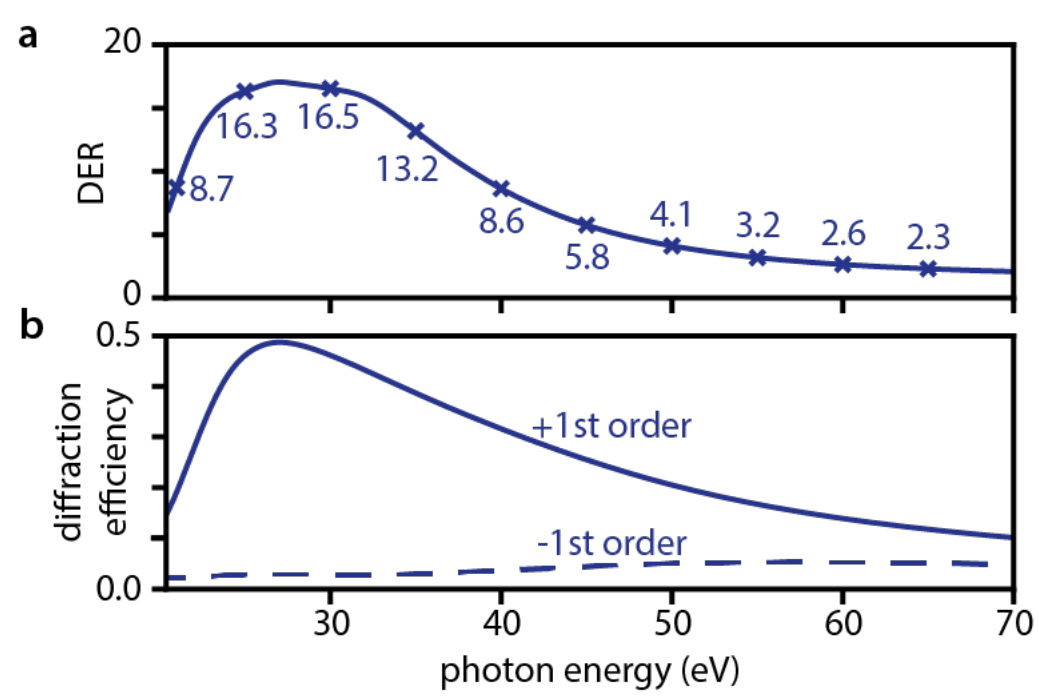

**Fig. 5: Broadband-optimized metagrating design.** **a** Diffraction efficiency ratio (DER) of an overetched metagrating (50 nm deeper holes). Crosses mark the photon energies between 20 and 65 eV in 5 eV steps and numbers give the corresponding DER. **b** Diffraction efficiency (relative to the full transmitted intensity) into the +1$^{st}$ diffraction order (solid blue curve) and into the -1$^{st}$ diffraction order (dashed blue curve) for an overetched metagrating.

## Conclusion

In EUV and attosecond spectroscopy, broadband, ultrashort pulses provide temporal resolution in a pump-probe experiment's interaction zone, while an angularly dispersive element after the interaction zone enables spectral resolution. Efficient dispersion is essential to resolve core-level transitions, enabling absolute energy calibration and element specificity, and ultrabroadband operation allows simultaneously observing multiple core-level transitions, revealing charge- and spin transfer on the fastest timescales [31,32].

Our experiments demonstrate that in-line transmissive blazed gratings - whether implemented as metagratings or as a height profile – meet and exceed these requirements: they suppress unwanted diffraction orders and enhance desired orders, offer high angular dispersion, broadband operation, all in a compact, simplified optical layout (in our case with a grating-detector distance of only 12.7 cm). When factoring in fabrication constraints, EUV metagratings can perform on par with sawtooth-blazed gratings fabricated via FIB milling, highlighting that even when used at the absolute wavelength limits, metasurface technology pushes what is possible.

This provides exiting opportunities: metasurfaces can achieve grating momenta beyond three times the reduced inverse unit cell size [33], enabling steep blaze angles and high angular dispersion. Whereas the samples sizes achievable by FIB milling are limited by the required write times (which can lead to ghosting if spatial beam or sample drifts during manufacturing are not cancelled), modern electron lithography systems can expose square-millimeter-sized metagratings in minutes, providing a route to large aperture samples. Although the absolute metagrating diffraction efficiency does not yet match that of the sawtooth grating, it can be improved by reducing the spacing between neighboring holes toward the design wavelength. Beyond existing devices, EUV metasurface technology opens avenues for combining focusing and spectral dispersion, orbital-angular momentum [34–36], and polarization-resolving devices.

**Materials and methods**

<u>Photon energy calibration</u>

To calibrate the absolute photon energy in measured diffraction spectra, we inserted a germanium filter into the beam path. Fig. S12 shows exemplary data and germanium's spectrally resolved transmission [37]. Both show a distinct transmission minimum at 32 eV.

<u>Metagrating fabrication</u>

The metagrating was fabricated (see Fig. S3) using electron beam lithography (EBL) followed by inductively coupled plasma reactive-ion etching (ICP-RIE). For EBL, the positive electron beam resist CSAR 62 (AR-P 6200.13) was spin-coated at 2800 rpm, resulting in a resist thickness of approximately 500 nm on the silicon membranes. The resist was exposed with a base dose of 65 $\mu C/cm^2$ (dose factor: 0.8 – 3.5, depending on the hole size) in a 30 kV Raith eLine plus system and developed in AR-P 600-546 for 90 seconds. Fig. S4 shows a SEM image of the metagrating pattern in the resist (before etching), while Fig. S5 compares the hole diameters extracted from Fig. S4 (red dots) and the target design hole diameters (blue dots). ICP-RIE was performed using an Oxford PlasmaPro Cobra 100 system. We etched the silicon membrane with a Pseudo-Bosch process using $SF_6$ (flow: 10 sccm) as the etching gas and $C_4F_8$ (flow: 12 sccm) as the passivation gas. The etch time was 170 seconds, at an ICP power of 400 W, with the sample glued on a Si <100> carrier wafer during etching. The left SEM image in Fig. S6 shows the fabricated metagrating after etching (final design).

Sawtooth grating fabrication

The sawtooth grating was produced via FIB milling (see Fig. S3) using a Scios2 HiVac Ga FIB with an ion milling current of 1 nA and a milling depth setting of 165 nm. The right SEM image in Fig. S6 shows the resulting sawtooth grating.

Metaatom simulations

To calculate the relative transmission phase and transmission (Fig. 4, S11) as a function of the hole diameter we performed RCWA simulations using GRCWA [38]. We considered holes in a 500 nm-thick silicon membrane with diameters ranging from 20 nm to 100 nm with a center-to-center distance (unit cell size) of 120 nm. We used the diameter-dependent hole depth ($h_{hole} = 4*d_{hole} - 70$ nm) measured from the FIB cross section in Fig. S2. This accounts for the reduced etch rate in smaller holes compared to larger ones (aspect ratio dependent etching (ARDE) [39,40]). The structure was excited with a normally incident plane wave, and the simulation was repeated for different wavelengths.

Grating simulations

We modeled diffraction efficiencies using FDTD simulations (tidy3d). Our metagrating simulation cell contained 12 unit cells (one grating period) and therefore 12 holes (with diameter-dependent depth) in a 500 nm-thick silicon layer. For the sawtooth grating, we modeled one grating period (1.5 µm) comprising a 220 nm-high sawtooth-shaped silicon layer on top of a 280 nm-thick silicon layer. To reproduce the fabricated structures, we adjusted the simulation structures accordingly. Perfectly matched layers were applied along the initial propagation axis and periodic boundary conditions otherwise. The excitation source was an s-polarized Gaussian pulse. Diffraction monitors were placed behind the gratings to extract the transmitted diffraction orders, which were normalized to the sum of the transmitted (not incident) light to match the experiment.

Modeling of phase based versus intensity based performance

To calculate the diffraction efficiencies into the $+1^{st}$ and $-1^{st}$ diffraction order for flat intensity and phase masks (results shown in Figs. 4c, d), we numerically modeled the complex transmission function of a one-dimensional grating using Fresnel integration [41]. The phase modulation was defined as a linear phase ramp with phase depth $\Delta\varphi \in [0, 2\pi]$ over one grating period ($\Lambda$ = 1.5 µm). The intensity modulation was implemented as a field amplitude variation proportional to the square root of a linear intensity profile, with $\Delta I \in [0, 1]$ over the same period. To mimic a realistic device, the grating was extended to 800 grating periods along the periodically changing direction and to 10 grating periods in the spatially uniform direction. The diffraction efficiency was normalized by the incident power.

**Code availability**

Exemplary diffraction efficiency modeling code used in this study is available at https://github.com/marcus-o/euv_metagrating_diffraction .

**Data availability**

The data that support the findings of this study are included in the manuscript and the supplementary materials and are archived at https://doi.org/10.5281/zenodo.20797219 .

**Acknowledgements**

We acknowledge Prof. Joachim R. Krenn from the University of Graz for providing access to their electron beam lithography system. This research was supported by the Scientific Service Units of ISTA through resources provided by the Nanofabrication Facility. The sawtooth grating was fabricated at the ELI-ALPS ERIC. The ELI ALPS project (GINOP-2.3.6-15-2015-00001) is supported by the European Union and co-financed by the European Regional Development Fund. The work of Z. P. and J. B. was supported by the Janos Bolyai Research Scholarship of the Hungarian Academy of Sciences (BO/00773/24 and BO/00686/24). M.O. acknowledges funding from the European Union (grant agreement 101076933 EUVORAM). The views and opinions expressed are, however, those of the author(s) only and do not necessarily reflect those of the European Union or the European Research Council Executive Agency. Neither the European Union nor the granting authority can be held responsible for them.

**Author Contributions**

A.K. and M.O. developed the project and conducted the numerical modeling. A.K. and T.R.C. designed the gratings. A.K., D.J.G. and A.H. developed and carried out the lithography process. A.K. and R.P. etched and imaged the samples. H.P. prepared the FIB cross-section for etch-depth analysis of the metagrating. Z.P., J.B., L.T., and P.D. fabricated and imaged the sawtooth grating. A.K., A.S.G, M.S., and M.O. designed and set up the laser frontend, spectral broadening, and HHG beamline. A.K., T.R.C., L.W., and M.O. conducted the experiment. A.K. and M.O. analyzed the experimental data and wrote the manuscript. All authors discussed the final version of the manuscript. M.O. acquired the project funding.

**Conflict of interest**

The authors declare no conflict of interest.

## Supplementary Information for

# Transmissive extreme ultraviolet metagrating

**Anna Kulter[1,*], Tiago Regio Crispim[1,2], Lorenz Weiss[1], Alexander Sagar Grossek[1], David J. Grafinger[1], Zsuzsanna Pápa[3], Judit Budai[3], Lázár Tóth[3], Péter Dombi[3,4], Rodolfo Previdi[5], Harald Plank[1], Andreas Hohenau[6], Martin Schultze[1], Marcus Ossiander[1,7,*]**

*[1] Graz University of Technology (Graz, 8010, Styria, Austria)*
*[2] Universidade do Minho (Braga, 4704-553, Norte, Portugal)*
*[3] ELI-ALPS, The Extreme Light Infrastructure ERIC (Szeged, 6728, Southern Great Plain, Hungary),*
*[4] HUN-REN Wigner Research Centre for Physics (Budapest, 1121, Central Hungary, Hungary)*
*[5] Institute of Science and Technology (Klosterneuburg, 3400, Lower Austria, Austria),*
*[6] Institute of Physics, University of Graz (Graz, 8010, Styria, Austria)*
*[7] Harvard University (Cambridge, Massachusetts, USA, 02138)*
*[*] anna.karner@tugraz.at, marcus.ossiander@tugraz.at*

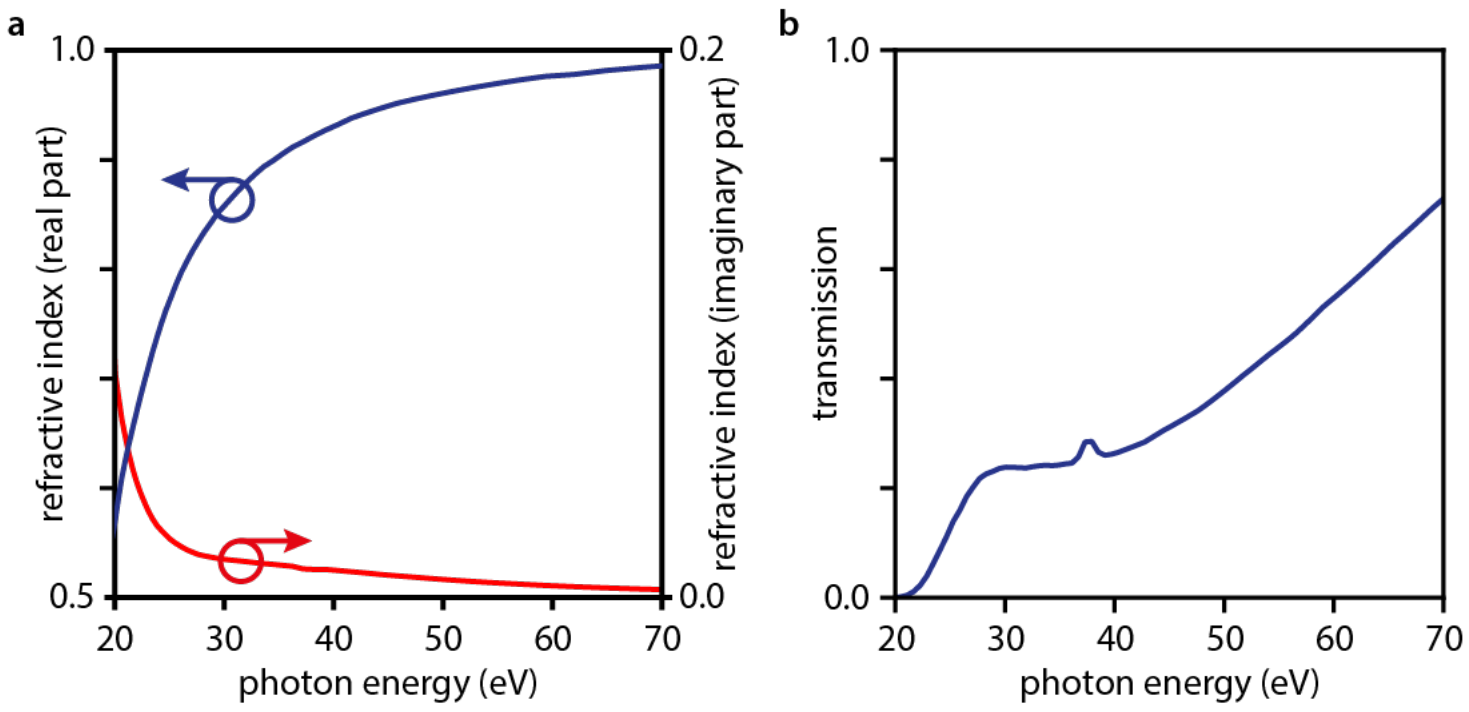

**Fig. S1 Refractive index and modeled transmission of a crystalline silicon membrane. a** Real and imaginary part of the refractive index of crystalline silicon in the extreme-ultraviolet spectrum (data from [20]). **b** Modeled transmission of a 500-nm thick silicon membrane.

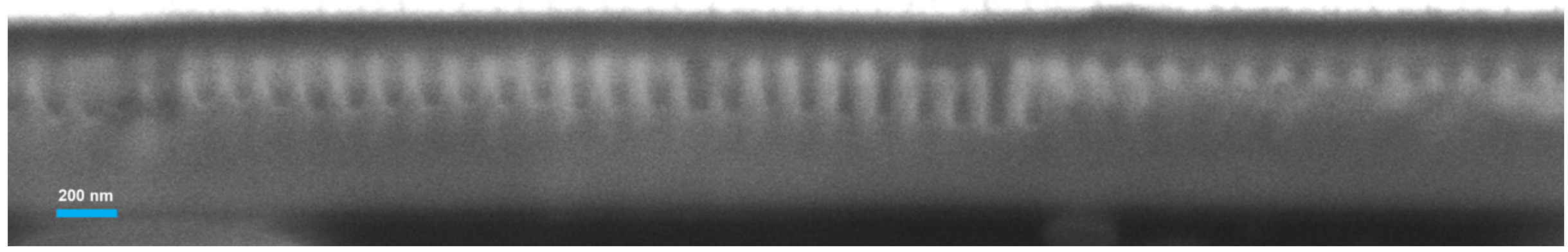

**Fig. S2 FIB cross-section from first etch trials.** Holes are not fully etched through after 2 minutes, with smaller holes etching more slowly than larger ones. The metagrating design accounts for this effect.

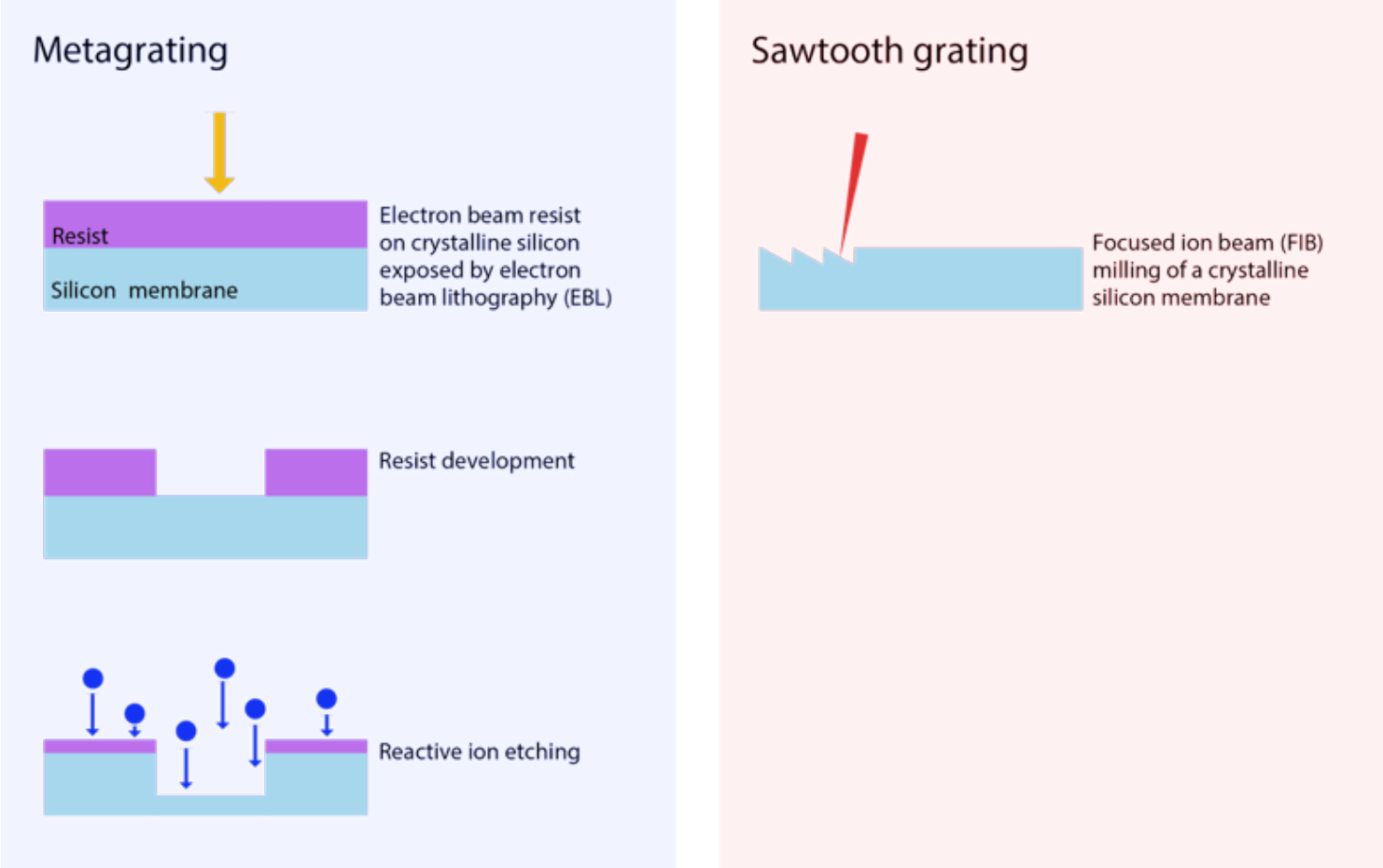

**Fig. S3 Fabrication workflow.** *Left:* Metagrating fabrication, starting with spin-coating a photoresist on the silicon membrane, followed by electron-beam exposure, resist development, and concluding with reactive ion etching. *Right:* Sawtooth grating fabrication; the pattern is directly milled into the membrane using a focused ion beam.[1]

1

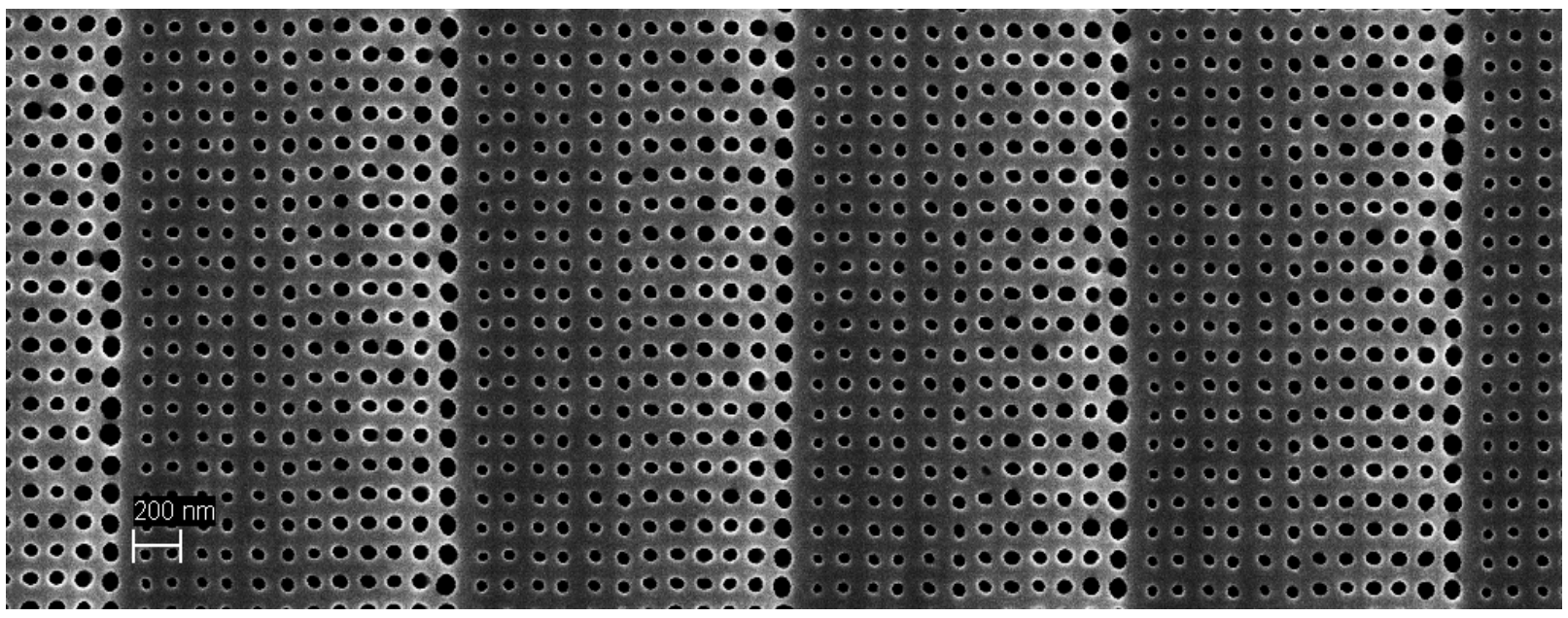

**Fig. S4 SEM image of a twin of the metagrating after lithography.** Metagrating pattern in the electron beam lithography resist.

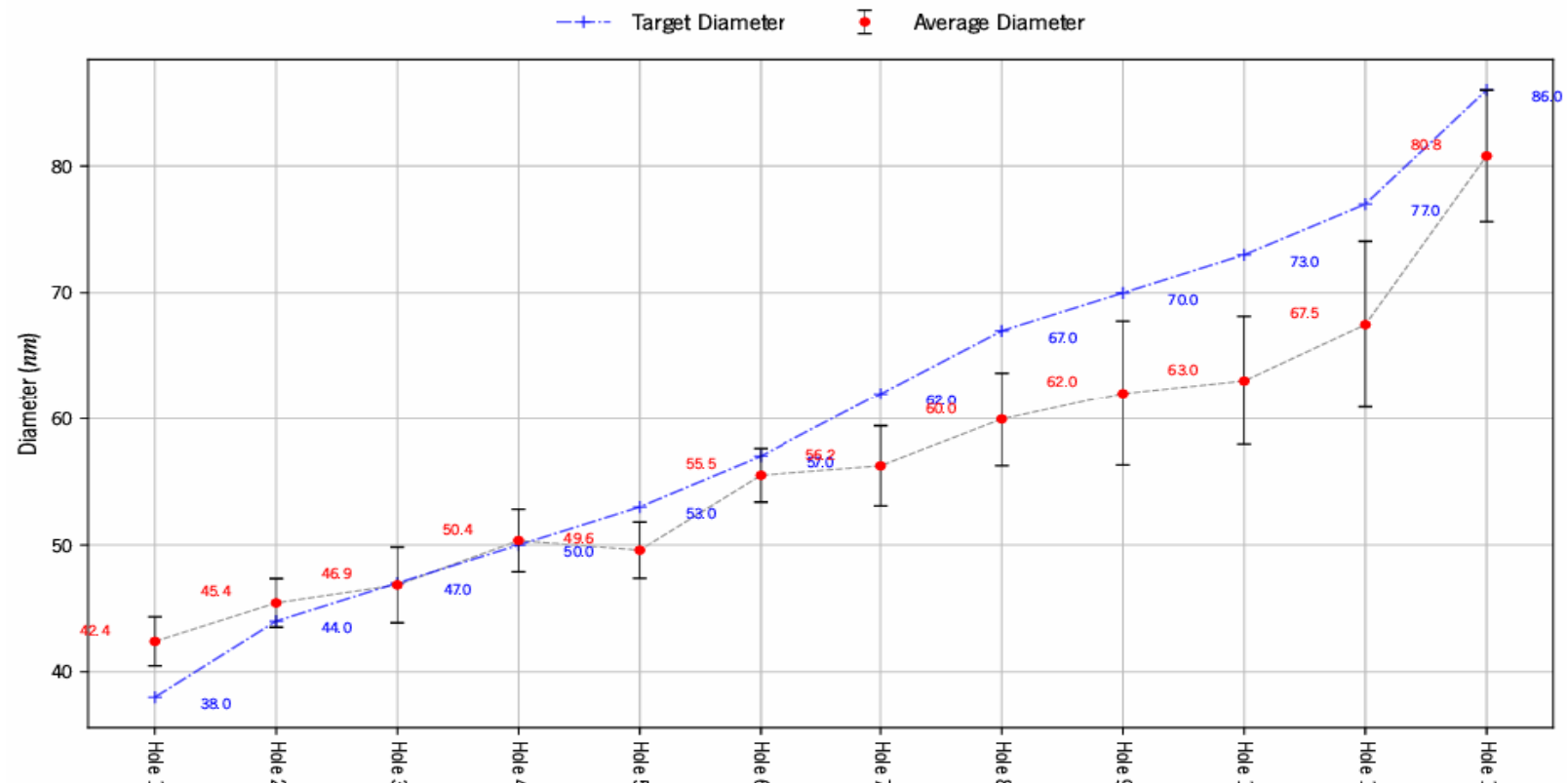

**Fig. S5 Design diameter versus actual hole diameter.** Comparison of the hole diameter in the metagrating lithography design (blue marks and line) with the diameter measured after lithography (red dots and gray line) from the image shown in Fig. S4.

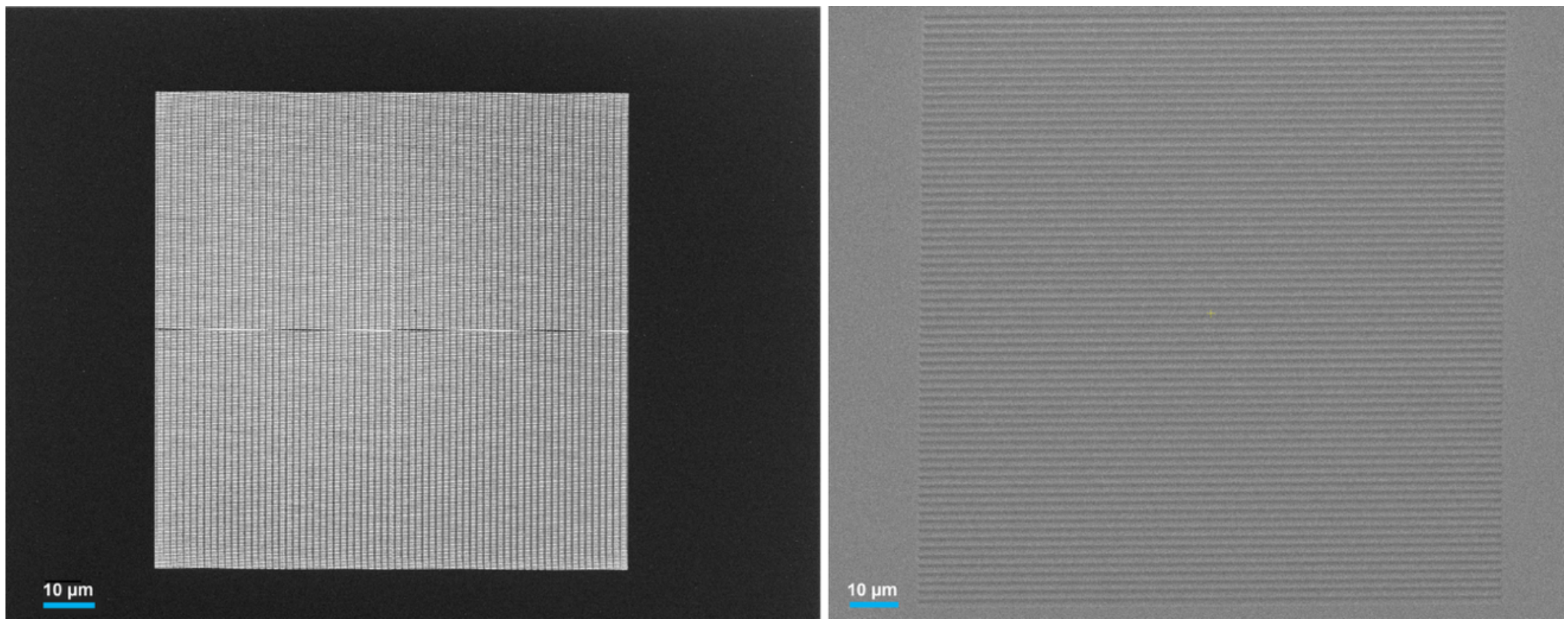

**Fig. S6 SEM images of the gratings.** Top view SEM images of the metagrating (left) and the sawtooth grating (right).

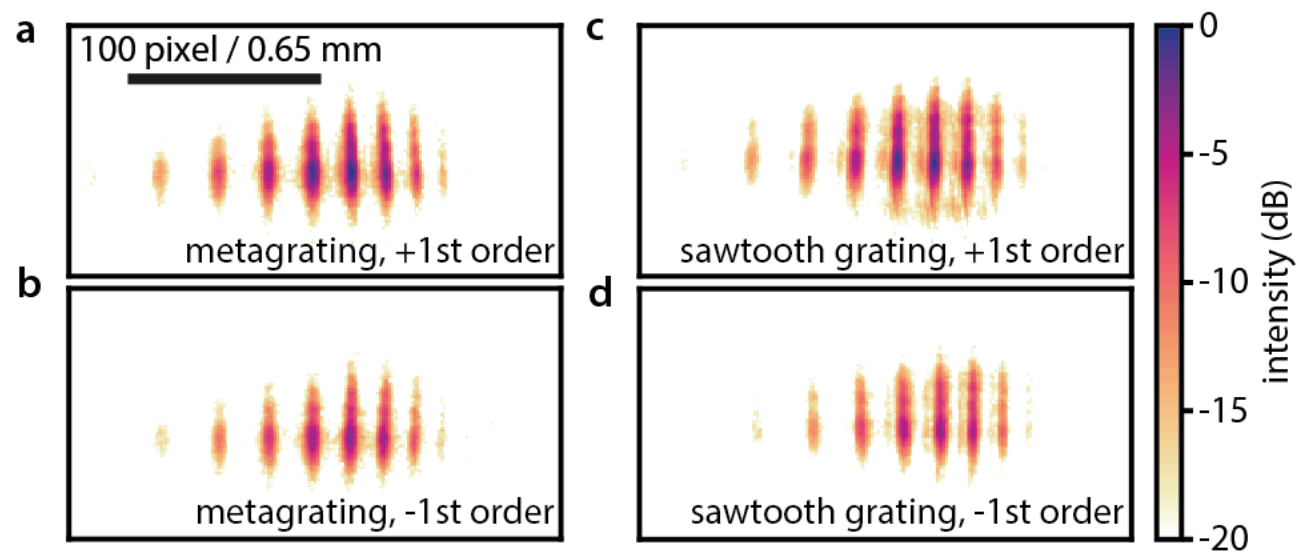

**Fig. S7: Diffraction patterns** of **a, (b)** +1$^{st}$, (-1$^{st}$) diffraction order generated by the metagrating and **c, (d)** +1$^{st}$, (-1$^{st}$) diffraction order generated by the sawtooth grating. Scale- and colorbar are valid for all panels.

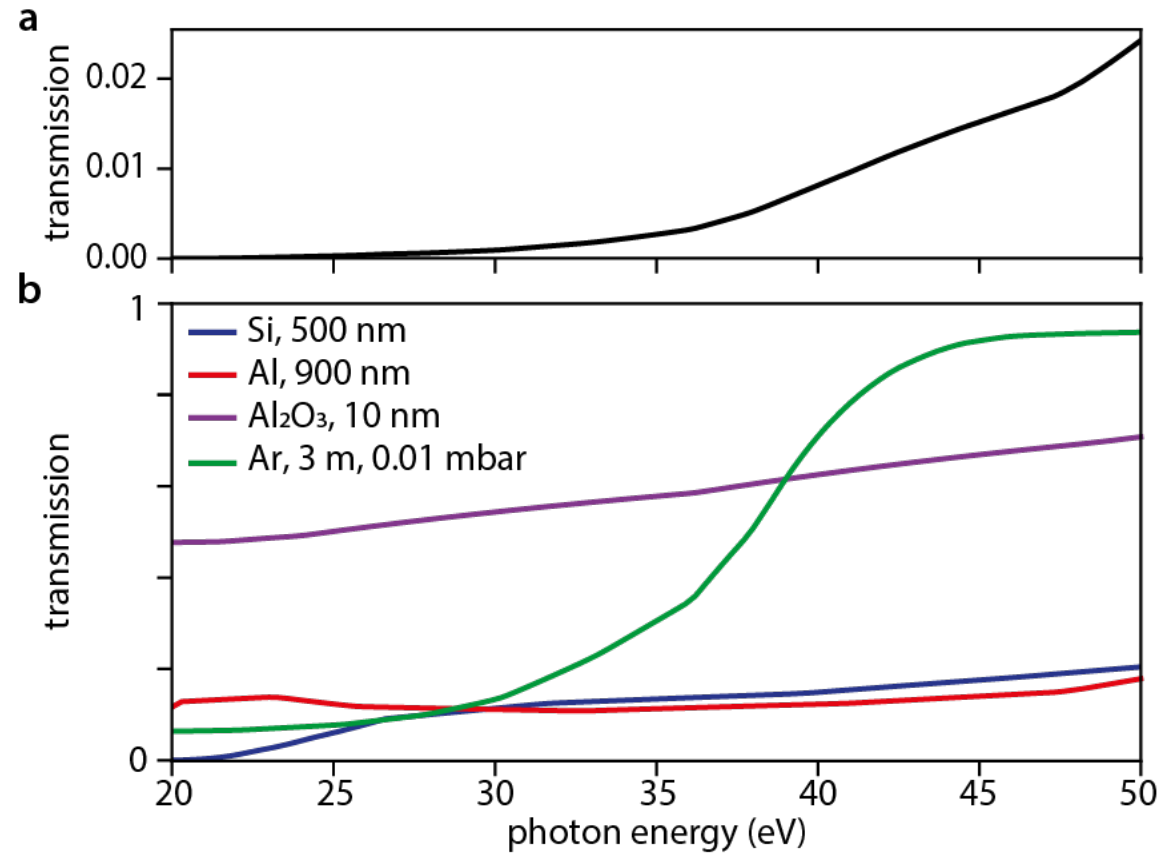

**Fig. S8 Transmission of silicon, aluminum, aluminum oxide and argon** [37]**.** **a** Cumulated transmission curves of 500 nm silicon, 900 nm aluminum, 10 nm alumina, and argon with a pressure of $10^{-2}$ mbar on a path length of 3 m. **b** Transmission of 500 nm silicon (Si, membrane thickness), 900 nm aluminum (Al, total filter thickness), 10 nm alumina ($Al_2O_3$, assumed thickness of total oxide layers on the filters) and argon (Ar) where we presumed a pressure of $10^{-2}$ mbar pressure in the chamber during high harmonic generation with argon and a path length of 3 m (beamline length from the HHG target to the camera).

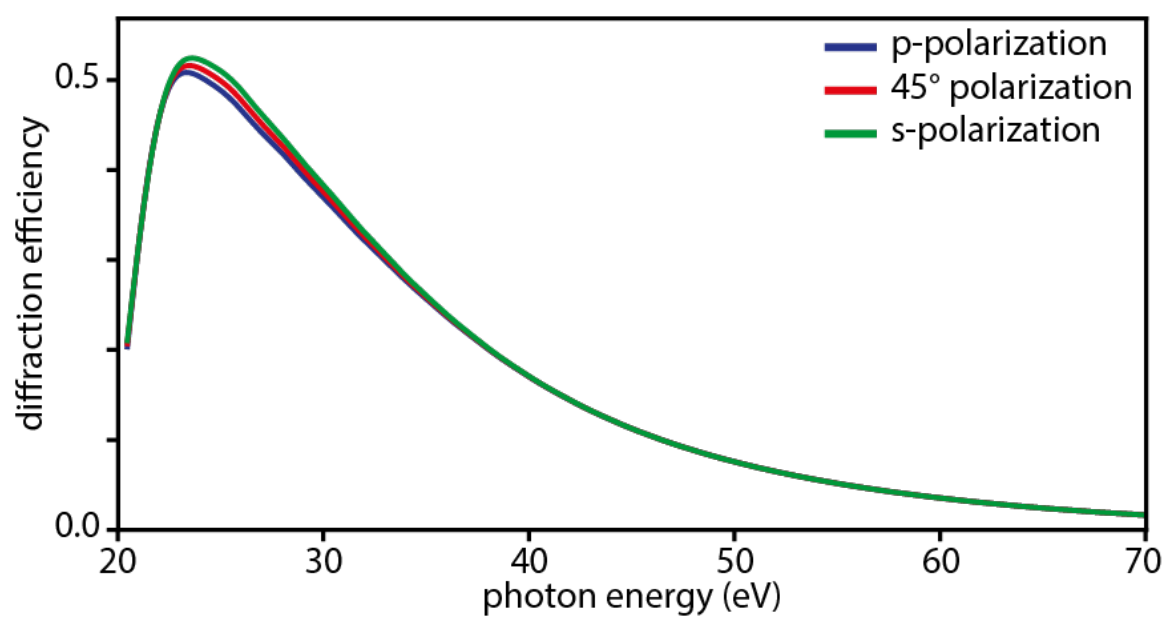

**Fig. S9 Polarization-dependent metagrating diffraction efficiency into the +1st order (modeled).** P-polarization describes an electric field of the generated +1$^{st}$ order parallel to the plane spanned by its propagation direction and the membrane's surface normal. S-polarization describes an electric field of the generated +1$^{st}$ order perpendicular to that plane. 45° polarization mixes both states.

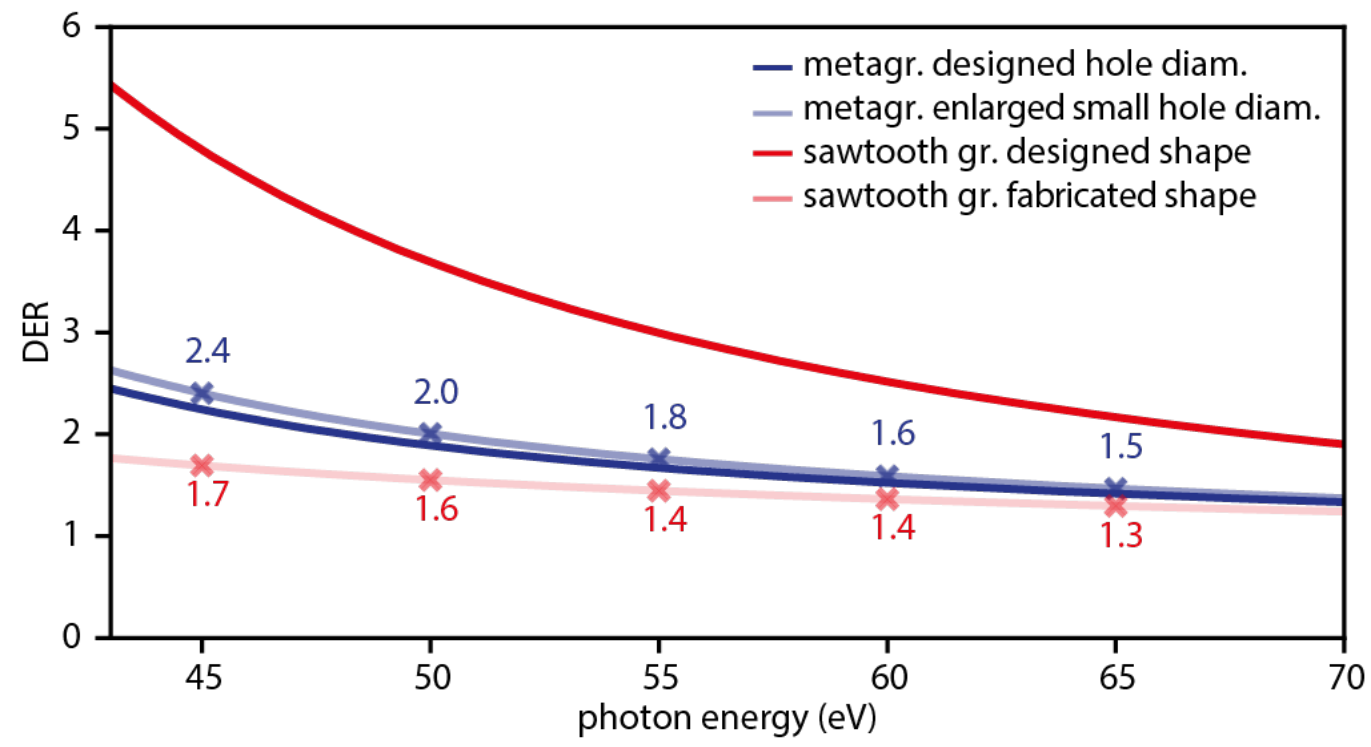

**Fig. S10 High-photon energy modeled diffraction efficiency ratio (DER)** for a true-to-design metagrating (dark blue curve), a metagrating in which the smallest holes are enlarged by 20 nm (light blue curve), for a true-to-design sawtooth grating (dark red curve) and the fabricated sawtooth grating (light red curve). Crosses mark the photon energies (45, 50, 55, 60, 65 eV) and numbers give DER values.

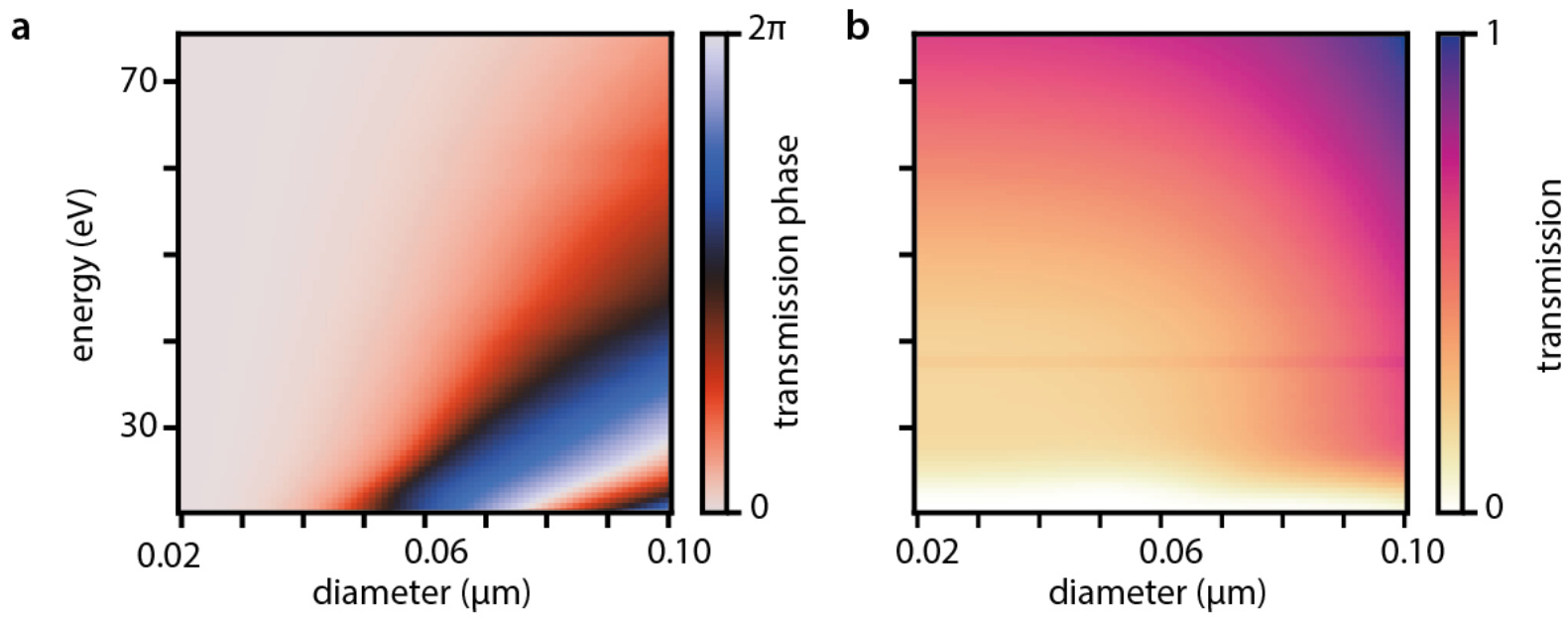

**Fig. S11 Spectrally resolved metaatom library.** **a** Simulated spectrally resolved and hole-diameter-dependent transmission phase and **b** transmission of periodically arranged holes in a silicon membrane.

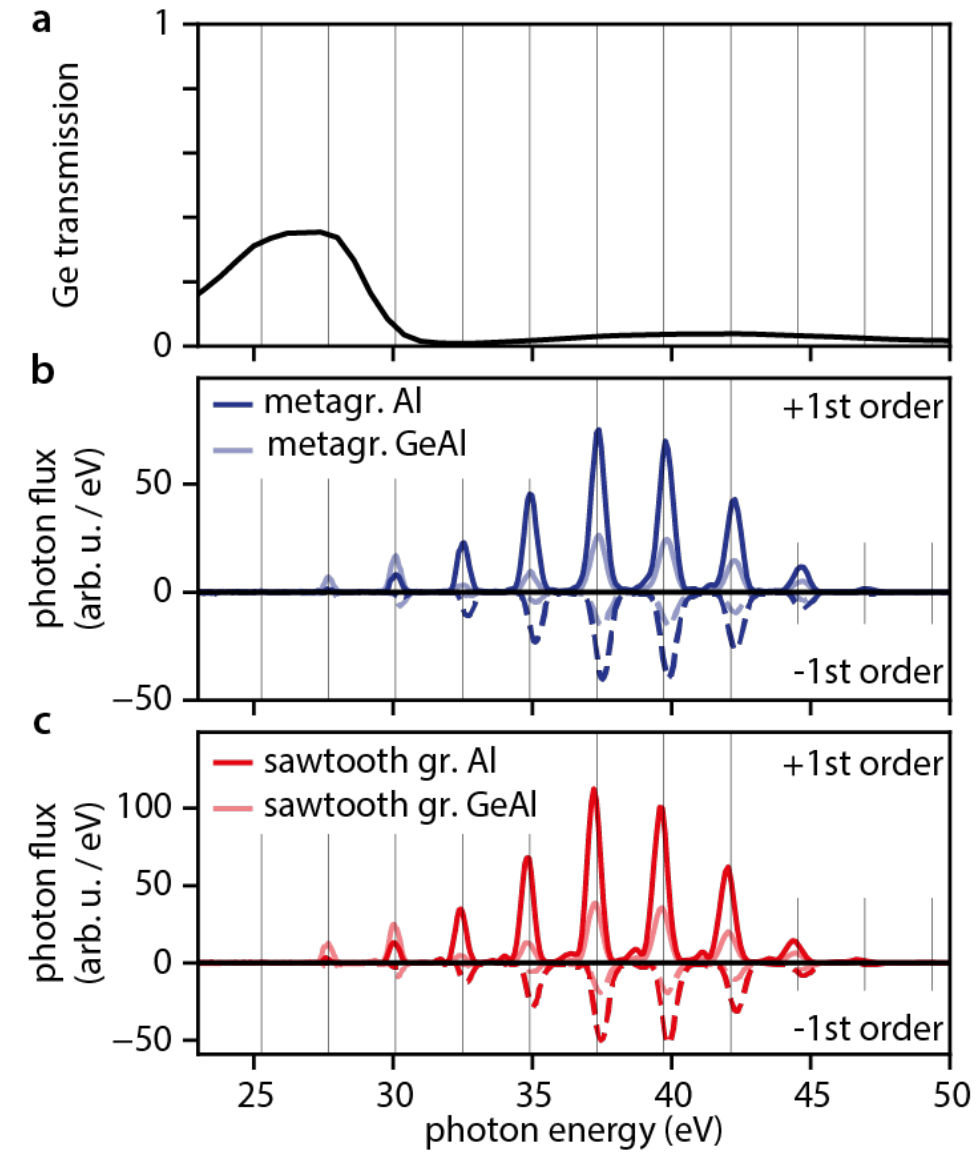

**Fig. S12 Experimental diffraction patterns after a germanium and an aluminum filter.**
**a** Germanium transmission [37]. Measured diffraction patterns of **b** the metagrating after an aluminum filter (blue line) and a germanium-aluminum filter (light blue line, scaled by 5) and **c** the sawtooth grating after an aluminum filter (red line) and a germanium-aluminum filter (light red line, scaled by 5). The positive y-axis denotes the $+1^{st}$ diffraction order, while the $-1^{st}$ diffraction order is shown on the negative y-axis (dashed lines). The gray lines indicate the harmonic photon energies of the driving laser. The comparison of the spectra after the different filters shows germanium's transmission minimum at 32 eV photon energy.